\documentclass[aps,prl,10pt,twocolumn,longbibliography,superscriptaddress]{revtex4-1}
\usepackage{amsmath}
\usepackage{amsfonts}
\usepackage{amssymb}
\usepackage{graphicx}
\usepackage{color}
\usepackage{hyperref}
\hypersetup{colorlinks=true,allcolors=blue}

\newcommand{\sbt}[1]{_\text{#1}} 
\newcommand{\spt}[1]{^\text{#1}} 
\newcommand{\KET}[1]{\vert #1\rangle} 
\newcommand{\ABS}[1]{\vert #1\vert} 

\begin{document}

\title{Two-Photon Excitation Sets Limit to Entangled Photon Pair Generation from Quantum Emitters}

\author{T. Seidelmann}
\email[Corresponding author: ]{tim.seidelmann@uni-bayreuth.de}
\affiliation{Lehrstuhl f{\"u}r Theoretische Physik III, Universit{\"a}t Bayreuth, 95440 Bayreuth, Germany}
\author{C. Schimpf}
\affiliation{Institute of Semiconductor and Solid State Physics, Johannes Kepler University Linz, 4040 Linz, Austria}
\author{T. K. Bracht}
\affiliation{Institut f{\"u}r Festk{\"o}rpertheorie, Universit{\"a}t M{\"u}nster, 48149 M{\"u}nster, Germany}
\author{M. Cosacchi}
\affiliation{Lehrstuhl f{\"u}r Theoretische Physik III, Universit{\"a}t Bayreuth, 95440 Bayreuth, Germany}
\author{A. Vagov}
\affiliation{Lehrstuhl f{\"u}r Theoretische Physik III, Universit{\"a}t Bayreuth, 95440 Bayreuth, Germany}
\author{A. Rastelli}
\affiliation{Institute of Semiconductor and Solid State Physics, Johannes Kepler University Linz, 4040 Linz, Austria}
\author{D. E. Reiter}
\altaffiliation[Current address: ]{Condensed Matter Theory, Department of Physics, TU Dortmund, 44221 Dortmund, Germany}
\affiliation{Institut f{\"u}r Festk{\"o}rpertheorie, Universit{\"a}t M{\"u}nster, 48149 M{\"u}nster, Germany}
\author{V. M. Axt}
\affiliation{Lehrstuhl f{\"u}r Theoretische Physik III, Universit{\"a}t Bayreuth, 95440 Bayreuth, Germany}

\begin{abstract}
Entangled photon pairs are key to many novel applications in quantum technologies. Semiconductor quantum dots can be used as sources of on-demand, highly entangled photons. The fidelity to a fixed maximally entangled state is limited by the excitonic fine-structure splitting. This work demonstrates that, even if this splitting is absent, the degree of entanglement cannot reach unity when the excitation pulse in a two-photon resonance scheme has a finite duration. The degradation of the entanglement has its origin in a dynamically induced splitting of the exciton states caused by the laser pulse itself. Hence, in the setting explored here, the excitation process limits the achievable concurrence for entangled photons generated in an optically excited four-level quantum emitter.
\end{abstract}

\maketitle

Entangled quantum states \cite{Horodecki:09,Orieux_entangled,entangled-photon2} inspired the development of applications in the fields of quantum cryptography \cite{Gisin:02,Lo_quantum_cryptography,Christian_Schimpf_Crypto_2021,Basset_quantum_key_2021}, quantum communication \cite{duan_quantum_comm,Huber_overview_2018}, or quantum information processing and computing \cite{pan:12,Bennett:00,Kuhn:16,Zeilinger_entangled}. The simplest realization of entangled qubits are entangled photon pairs. These are attractive due to their robustness against environmental decoherence \cite{Orieux_entangled}. In the past decades, quantum dots (QDs) have emerged as a versatile platform for the generation of polarization-entangled photon pairs in experiment \cite{Benson_2000_QD_cav_device,Stevenson2006,Young_2006,Hafenbrak,Muller_2009,dousse:10,Biexc_FSS_electrical_control_Bennett,stevenson:2012,entangled-photon1,Trotta_highly_entangled_2014,winik:2017,huber2017,Huber_PRL_2018,Bounouar18,Wang_2019,Liu2019,Fognini_2019,
Hopfmann_2021} and theory  \cite{BiexcCasc_Carmele,Jahnke2012,heinze17,Different-Concurrences:18,Seidelmann2019,Phon_enhanced_entanglement,EdV,Troiani2006} as well as for time-bin entangled photon pairs \cite{jayakumar2014time,huber2016coherence,gines2021time}.

\begin{figure}[t]
\centering
\includegraphics[width=\columnwidth]{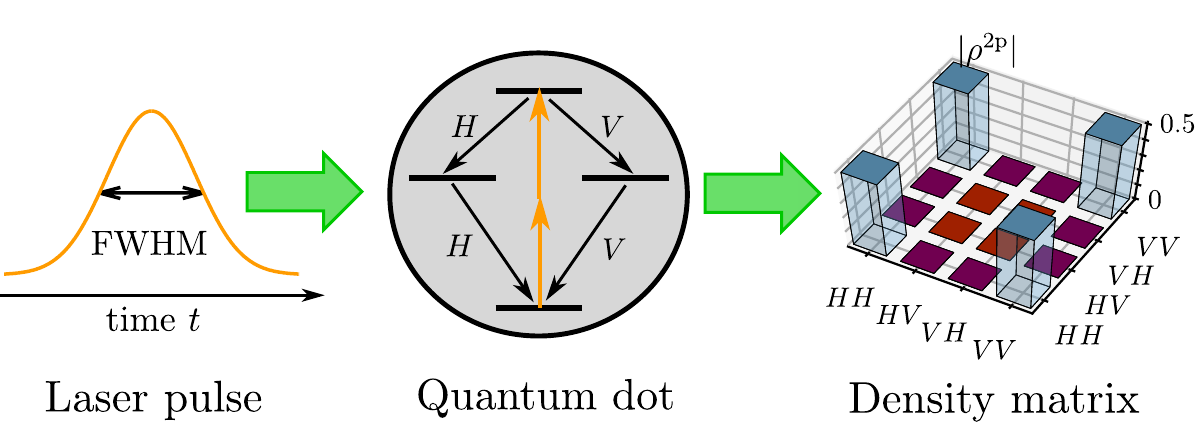}
\caption{
A Gaussian pulse (left) with finite full width at half maximum (FWHM)  excites a quantum dot. In the two-photon resonant excitation scheme, the central frequency of the laser pulse matches half of the ground state-to-biexciton transition energy. The excited quantum dot decays radiatively, following the diamond-shaped cascade. Ideally, either a pair of horizontally ($H$) or vertically ($V$) polarized photons is emitted (center), resulting in a maximally entangled state $\KET{\Phi_+}$ described by the ideal two-photon density matrix $\rho\spt{2p}$ (right).
}
\label{fig:system}
\end{figure}

Their generation relies on the biexciton-exciton cascade. After the biexciton is prepared, it decays by one of two paths, cf., Fig.~\ref{fig:system} middle panel, ideally emitting either a pair of horizontally ($H$) or vertically ($V$) polarized photons. In the ideal case of zero fine-structure splitting, the information of the chosen path (which-path information) is missing, and the  resulting two-photon state
\begin{equation}
\label{eq:max_ent_state}
\KET{\Phi_+} = \frac{1}{\sqrt{2}}\left( \KET{HH} + \KET{VV} \right)
\end{equation}
is a maximally entangled Bell state. However, in reality, the generated state deviates from the perfect Bell state. This deviation can be quantified by the fidelity $\cal F$ \cite{Jozsa_fidelity} or the concurrence $C$ \cite{Wootters1998}, defined such that only ${\cal F}=1$ or $C=1$ corresponds to a maximally entangled Bell state.

The fine-structure splitting (FSS), i.e., an energy difference between the two exciton levels, is a major obstacle for generating perfectly entangled states.  By breaking the symmetry of the system, it introduces which-path information during the photon generation \cite{Hudson2007}. Several methods were developed to suppress the FSS \cite{Muller_2009,Biexc_FSS_electrical_control_Bennett,huber2017,Huber_overview_2018}, resulting in entangled states with higher concurrence. From the theory side, it was predicted that an initially prepared biexciton yields a maximally entangled state in the case of vanishing FSS \cite{BiexcCasc_Carmele,Seidelmann2019}, even when cavity and radiative losses as well as phonons are considered. 

However, the influence of the preparation process is still under debate. Early experimental proof-of-principle studies often employed far off-resonant excitation schemes, where carriers were excited in higher QD states or the wetting layer, which subsequently relaxed into the biexciton state \cite{Young_2006,Hafenbrak,Muller_2009,dousse:10,Biexc_FSS_electrical_control_Bennett}. Recent state-of-the-art experiments rely on a coherent two-photon resonant excitation (TPE) scheme with typical pulse durations on the order of 10 ps and light with linear polarization in the QD growth plane, and demonstrate very high degrees of entanglement \cite{winik:2017,huber2017,Huber_PRL_2018,Wang_2019,Liu2019,Hopfmann_2021,
Christian_Schimpf_Crypto_2021,Christian_Schimpf_strongly_ent_2021}, because of the strongly reduced reexcitation probability \cite{schweickert2018,hanschke2018}. Nevertheless, perfectly entangled photons have not yet been observed, and the question remains whether TPE can still be a source of entanglement degradation.

In this letter, we show that the TPE scheme employed to create the biexciton sets a limit to the obtainable concurrence. To demonstrate this, we present numerical simulations alongside analytical calculations giving a dependence of the maximally achievable concurrence as a function of the full width at half maximum (FWHM) of the exciting Gaussian laser pulse. We provide a clear physical picture how the laser introduces which-path information that significantly reduces the achievable degree of entanglement.

We model the QD as a four-level system, cf., Fig.~\ref{fig:system}, consisting of the ground state $\KET{G}$, excitons with horizontal and vertical transition dipoles $\KET{X\sbt{H}}$ and $\KET{X\sbt{V}}$, and the biexciton $\KET{B}$. The FSS between the two exciton states, which is typically on the order of a few $\mu$eV, is denoted as $\delta$ and we assume that the energy of $\KET{X\sbt{H}}$ is always higher than the one of $\KET{X\sbt{V}}$. $E\sbt{B}$ is the biexciton binding energy, i.e., the difference between the energy of two uncorrelated excitons and the biexciton. The biexciton (an exciton) decays with a characteristic rate $\gamma\sbt{B}$ ($\gamma\sbt{X}$) into an exciton (the ground) state. For the numerical calculations, we use parameters for typical GaAs QDs given in Tab.~\ref{tab:Fixed_Parameters}.

\begin{table}
\centering
\caption{Quantum dot parameters used in the calculations.}
\label{tab:Fixed_Parameters}
\begin{ruledtabular}
\begin{tabular}{l c c}
Parameter & & Value\\
\hline
Biexciton binding energy & $E\sbt{B}$ & 4~meV \\
Exciton-laser detuning & $\Delta\sbt{XL}$ & $E\sbt{B}/2=2$~meV \\
Radiative decay rate exciton & $\gamma\sbt{X}$ & 0.005~$\mathrm{ps^{-1}}$\\
Radiative decay rate biexciton & $\gamma\sbt{B}$ & $2\gamma\sbt{X}=0.010$~$\mathrm{ps^{-1}}$
\end{tabular}
\end{ruledtabular}
\end{table}

In experiments, the concurrence is obtained from the two-photon density matrix which is reconstructed employing quantum state tomography \cite{QuantumStateTomography}. This scheme relies on measuring polarization-resolved two-time correlation functions, where the measurement represents a statistical average over both time arguments - the real time of the first detection event and the delay time until a subsequent second one. In principle, one can restrict the averaging intervals to narrow time windows, which corresponds to selecting different subsets of photon pairs \cite{Different-Concurrences:18}. Such time filtering can be used to alleviate dephasing effects, but at the cost of a reduced photon generation. Here,
the two-photon density matrix $\rho\spt{2p}$ [in the basis $\lbrace \KET{HH}, \KET{HV}, \KET{VH},\KET{VV}\rbrace$] is calculated based on time-integrated correlation functions, where we average over all possible real and delay times. Details of the theoretical model, the two-photon density matrix, and the concurrence can be found in Supplemental Material \cite{supp}.\nocite{Lindblad:1976,LossDivincenzo_SW,Winkler,multi-time,Seidelmann_QUTE_2020}

It is instructive to briefly recapitulate the concurrence predicted for the initial value problem where one assumes an initially prepared biexciton. In the ideal situation of a vanishing FSS, the energy structure is completely symmetric, and the resulting two-photon state is maximally entangled. But, if the FSS is finite, the two emission paths in the biexciton-exciton cascade can be distinguished, and emitted photons with opposite polarization exhibit slightly different energies. Thus, a finite FSS introduces which-path information into the system. In the two-photon density matrix this which-path information manifests itself in a reduced coherence between the states $\KET{HH}$ and $\KET{VV}$. The corresponding concurrence can be calculated analytically, cf., Supplemental Material \cite{supp}, yielding  
\begin{equation}
\label{eq:con_initial}
C_0(\gamma\sbt{X},\delta) = \frac{1}{\sqrt{1+\left( \frac{\delta}{\hbar\gamma\sbt{X}} \right)^2}}
\end{equation} 
It depends solely on the ratio between the FSS $\delta$ and the decay rate $\gamma\sbt{X}$ of the exciton states. We stress that in the case of $\delta=0$, the concurrence is unity for an initially prepared biexciton, i.e., the maximally entangled Bell state is created. In particular, it was shown that the concurrence in this case is robust against dephasing processes \cite{Seidelmann2019,BiexcCasc_Carmele,Different-Concurrences:18}.

For a finite FSS, $C_0$ reflects the integration over a time-dependent phase oscillation of the exciton coherence during the measurement process \cite{Hudson2007}. Because of this oscillating phase, the two-photon state associated with one possible cascaded decay is
\begin{equation}
\KET{\Phi_\tau} = \frac{1}{\sqrt{2}}\left( \KET{HH} + e^{i\frac{\delta}{\hbar}\tau} \KET{VV} \right)
\end{equation}
where $\tau$ is the delay time between the first (biexciton) and second (exciton) photon emission event. Averaging over all possible realizations with different delay times results in a mixed state with a reduced coherence. Note that the concurrence depends on the averaging window for the delay time in the experiment. Selecting only photon pairs in a small delay-time window results in a higher concurrence \cite{winik:2017,Fognini_2019}, but one has to discard the majority of emission events. Furthermore, when the QD is embedded inside a cavity, even in the limit of a vanishing averaging window, the concurrence does not reach unity \cite{Different-Concurrences:18}.

\begin{figure}[t]
\centering
\includegraphics[width=\columnwidth]{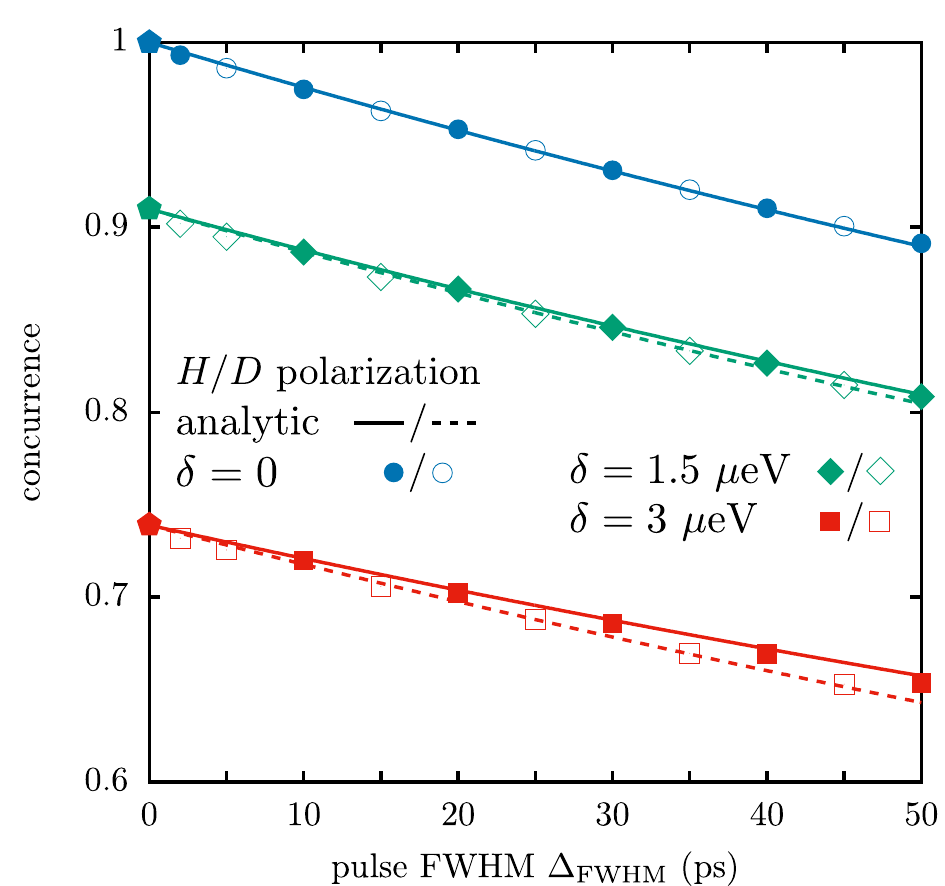}
\caption{
Concurrence in dependence of the pulse duration, characterized by its FWHM, for two laser polarizations [horizontal ($H$) filled symbols and diagonal ($D$) open symbols] and three fine-structure splittings $\delta=0$ (blue circles), 1.5 $\mathrm{\mu}$eV (green diamonds), and 3 $\mathrm{\mu}$eV (red squares). The symbols represent numerical results and lines are the analytic expression Eq.~\eqref{eq:C_full_estimate} ($H$ polarization: solid line; $D$ polarization: dashed line). For $\delta=0$, the results for $H$ and $D$ polarization are exactly the same. Data points at $\Delta_\text{FWHM}=0$ (pentagons) represent calculations with an initially prepared biexciton without optical excitation.
}
\label{fig:no_phonons}
\end{figure}

Having seen that in the limit of vanishing FSS, an initially prepared biexciton yields unity concurrence, we now consider the impact of the excitation process in the TPE scheme. Here, the biexciton is excited by a laser pulse in resonance with the two-photon biexciton transition. In the simulations, we assume a Gaussian pulse with envelope 
\begin{equation}
\Omega(t) = \sqrt{\frac{4\ln (2)}{\pi}}\frac{\Theta}{\Delta_\text{FWHM}}\exp\left[ -4\ln(2)\left(\frac{t}{\Delta_\text{FWHM}}\right)^2 \right]
\end{equation}
where the FWHM of the laser pulse $\Delta_\text{FWHM}$ is the central quantity of interest. Note that the width of the corresponding laser intensity $I(t)\propto\Omega^2(t)$ is characterized by  $\Delta_\text{FWHM}/\sqrt{2}$. To achieve a two-photon $\pi$ pulse, the pulse area $\Theta$ is determined numerically, such that the maximal biexciton occupation is obtained. The optimal value is roughly
\begin{equation}
\Theta \approx \sqrt{\frac{E\sbt{B}\,\Delta_\text{FWHM}}{\hbar\sqrt{2\pi\ln (2)}}}\,\pi
\end{equation}
cf., Supplemental Material \cite{supp}. Note that the concurrence is not sensitive to the initial occupation of the biexciton - it can reach unity also for a partially occupied biexciton state.

We start with the case of vanishing FSS, i.e., $\delta=0$, and consider two different linear laser polarizations: (i) horizontal, i.e., the laser polarization coincides with the polarization $H$, and (ii) diagonal, i.e., the laser polarization has equal components in $H$ and $V$. As shown in Fig.~\ref{fig:no_phonons} (blue symbols), both laser polarizations result in the same concurrence. In contrast to the initial value problem without optical excitation, i.e., data points at $\Delta_\text{FWHM}=0$, the concurrence drops significantly with increasing FWHM. For a typical pulse length of 10~ps \cite{Huber_PRL_2018,Liu2019,Christian_Schimpf_Crypto_2021}, the concurrence decreases to $C\approx 0.975$. Using a dressed state picture, we can derive (see Supplemental Material \cite{supp}) the analytic expression
\begin{subequations}
\label{eq:Con_estimate_FSS_0}
\begin{equation}
C \approx 1-2f(\gamma\sbt{B},\Delta_\text{FWHM})
\end{equation}
\begin{equation}
\label{eq:def_f}
f(\gamma\sbt{B},\Delta_\text{FWHM}) = \frac{\gamma\sbt{B}\Delta_\text{FWHM}}{8}\,\exp\left[-\frac{\gamma\sbt{B}\Delta_\text{FWHM}}{4} \right]
\end{equation}
\end{subequations}
which describes the drop of concurrence with increasing FWHM well. The obtained result depends solely on the product of the FWHM and the biexciton decay rate, which can be interpreted as a measure for the number of biexciton photons that are emitted during the pulse. We stress that this drop does not originate from an increasing reexcitation probability, cf., Supplemental Material \cite{supp}, i.e., the creation of additional photons is negligible.

\begin{figure}[t]
\centering
\includegraphics[width=\columnwidth]{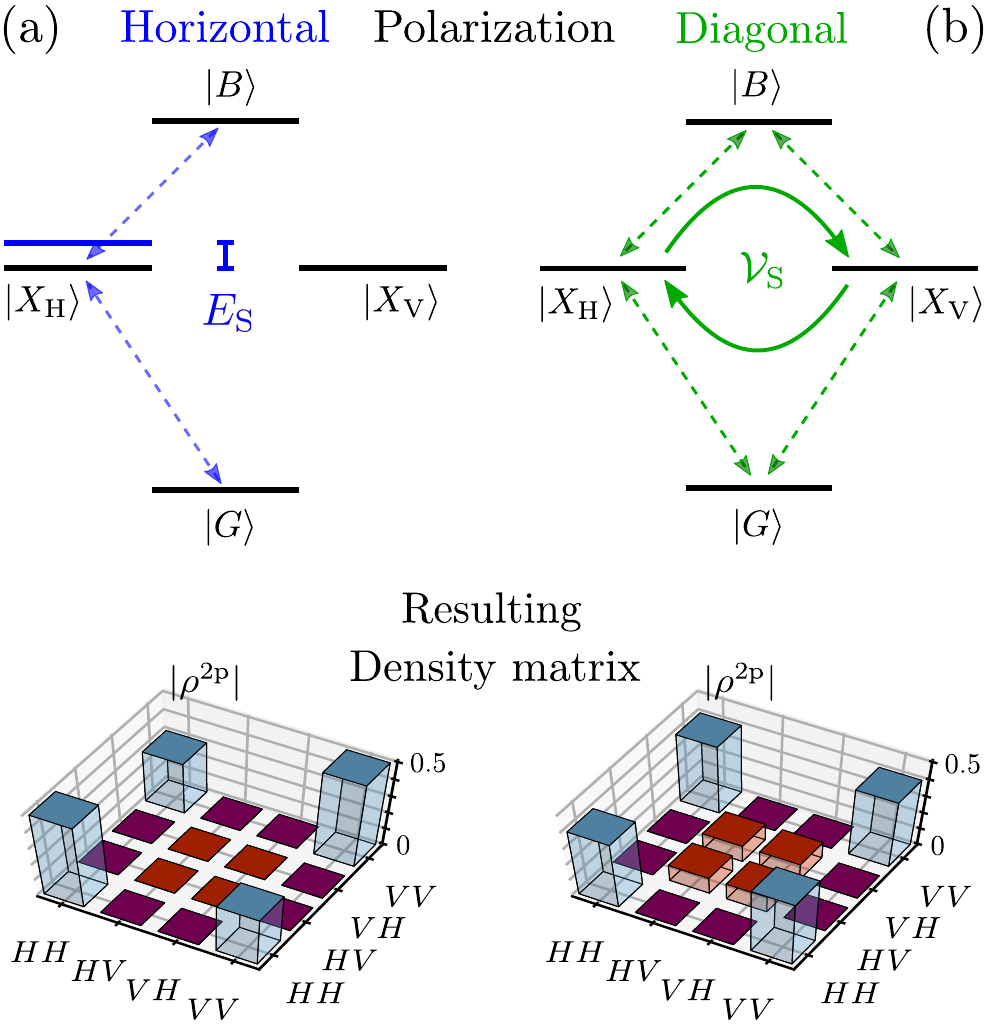}
\caption{
Top: sketches of the four-level system with the laser-induced effects for (a) horizontal polarization and (b) diagonal polarization. Black lines are the unperturbed quantum dot states. Optical transitions are indicated by dashed arrows. The laser effect for horizontally polarized excitation can be interpreted as an AC Stark shift yielding an energetic splitting $E\sbt{S}$. For a diagonal polarization, the laser-induced interaction can be interpreted as a coupling between the excitons with coupling strength $\mathcal{V}\sbt{S}$. Bottom: examples of resulting density matrices for the two polarizations.
}
\label{fig:dressing_effects}
\end{figure}

To understand this drop in concurrence, we start with considering the excitation with horizontal ($H$) polarization. We recall that a reduced degree of entanglement is associated with which-path information. The only source of the latter in the TPE scheme is the laser pulse, which makes the level configuration (dynamically) asymmetric as illustrated in Fig.~\ref{fig:dressing_effects}(a). Only $\KET{X\sbt{H}}$ interacts with the laser pulse, while $\KET{X\sbt{V}}$ remains unchanged. The interaction with this laser pulse introduces an AC Stark shift for $\KET{X\sbt{H}}$, resulting in a finite energy shift $E\sbt{S}\propto\Omega^2(t)/E\sbt{B}$. For the two-photon $\pi$ pulse, this energy shift is on the order of
\begin{equation}
E\sbt{S}\sim\frac{\hbar\pi}{\Delta_\text{FWHM}}
\end{equation}
cf., Supplemental Material \cite{supp}.
Hence, during the pulse, the two exciton states are split by $E\sbt{S}$.
As in the case of a fixed FSS, the dynamic splitting $E\sbt{S}$ gives rise to phase oscillations of the exciton coherence until the pulse is gone. Thus, on the timescale of the pulse, which-path information is introduced by the TPE scheme itself. Note that for a FWHM of 10 ps, $E\sbt{S}\sim 200$ $\mu$eV is  one order of magnitude larger than typical FSSs. With increasing pulse length the dynamically induced splitting $E\sbt{S}$ becomes smaller but persists in a longer time window. Thus, more emitted photon pairs are affected by it, resulting in a monotonic drop of the concurrence with rising FWHM. It is instructive to consider the effect of this which-path information on the two-photon density matrix as shown in Fig.~\ref{fig:dressing_effects}(a), bottom row. Similar to what occurs for a finite FSS, we observe a reduced coherence $\rho\spt{2p}_{HH,VV}$.

Because the reduced concurrence has been traced back to an asymmetry during the pulse, one could naively assume that using a diagonally polarized laser might result in a maximally entangled state. But numerical calculations with a diagonal polarization yield exactly the same degree of entanglement, cf., Fig.~\ref{fig:no_phonons}. Of course, when the FSS is absent, all orthogonal bases, constructed as linear combinations of the horizontally and vertically polarized exciton state, are equivalent. Consequently, all linear laser polarizations yield the same entanglement. Clearly, this applies to the basis states $\KET{X\sbt{D/A}} := (\KET{X\sbt{H}}\pm\KET{X\sbt{V}})/\sqrt{2}$, for which the system becomes identical to the one with the horizontal laser polarization.

It is nevertheless instructive to look at the interpretation of the drop in concurrence for a diagonally polarized excitation. During the laser pulse, the action of the pulse can be described as a full-amplitude (coherent) oscillation between the two different exciton states (cf., Supplemental Material \cite{supp}). Thus, if the QD decays into an exciton state via the emission of a biexciton photon already during the pulse duration, the pulse introduces an effective coupling to the other exciton. In Fig.~\ref{fig:dressing_effects}(b) this coupling is denoted as $\mathcal{V}\sbt{S}$. Therefore, during the pulse duration, the exciton state can change and the subsequently emitted exciton photon can have a different polarization than the prior biexciton photon. Consequently, the interaction with the laser enables the creation of two-photon states $\KET{HV}$ and $\KET{VH}$, which represent a deviation from the maximally entangled state $\KET{\Phi_+}$. This interpretation is confirmed by looking at the two-photon density matrix in Fig.~\ref{fig:dressing_effects}(b), where the elements $\rho\spt{2p}_{HV,HV}$ and $\rho\spt{2p}_{VH,VH}$ become finite, causing a reduced degree of entanglement.
Note that for diagonal polarization the occupations  $\rho\spt{2p}_{HH,HH}$ and  $\rho\spt{2p}_{VV,VV}$ have the same value as the coherence  $\ABS{\rho\spt{2p}_{HH,VV}}$ in contrast to the excitation with horizontal polarization where the relative amplitude of this coherence is reduced.

Thus, a diagonal laser polarization corresponds to describing the same effect in a different basis or picture. This is similar to the FSS stemming from the anisotropic exchange interaction, which can also be discussed as an interaction between circularly polarized excitons or an energetic splitting $\delta$ between linearly polarized ones. In the TPE scheme, the electromagnetic field plays the role of a tunable exchange interaction.

Finally, we analyze the combined effect of FSS and laser-induced splitting. Figure~\ref{fig:no_phonons} shows the concurrence obtained for two finite FSSs $\delta=1.5$ $\mathrm{\mu}$eV (green) and 3 $\mathrm{\mu}$eV (red) for a horizontal ($H$: filled symbols) and diagonal ($D$: open symbols) laser polarization. We remined the reader that in the limit of a vanishing pulse duration, i.e., $\Delta_\text{FWHM}\rightarrow 0$, the concurrence is given by $C_0(\gamma\sbt{X},\delta)$. Starting from this value, the concurrence drops with rising FWHM due to the laser-induced splitting. Depending on the laser polarization, this can be observed either as an additional loss of coherence (horizontal polarization) or an increase of detrimental photon states (diagonal polarization) in the two-photon density matrix, cf., Supplemental Material \cite{supp}. Because the effect of the FSS on the TPE scheme is negligible for $\delta \ll E\sbt{B}$, we find only a marginal difference between the two polarizations. 

In this case, the drop in concurrence is well described by the analytic expression
\begin{subequations}
\label{eq:C_full_estimate}
\begin{eqnarray}
C &\approx& C_0(\gamma\sbt{X},\delta) \left\lbrace 1-f(\gamma\sbt{B},\Delta_\text{FWHM}) \left[ 1+g(\alpha_H) \right] \right\rbrace \notag \\ 
&&- f(\gamma\sbt{B},\Delta_\text{FWHM}) \left[ 1-g(\alpha_H) \right] \\
g(\alpha_H) &=& \left( 1-2\alpha_H^2 \right)^2
\end{eqnarray}
\end{subequations}
where $C_0$ [defined in Eq.~\eqref{eq:con_initial}] represents the concurrence associated with an initially prepared biexciton, while $f$ [defined in Eq.~\eqref{eq:def_f}] and $g$ capture the influence of the pulse duration and the laser polarization, respectively.
The parameter $\alpha_H$ describes the component of the laser polarization in the $H$ direction such that the horizontal (diagonal) laser polarization corresponds to $\alpha_H = 1$ ($\alpha_H = 1/\sqrt{2}$). Note that for $\delta=0$, this expression reduces to Eq.~\eqref{eq:Con_estimate_FSS_0} and becomes independent of the laser polarization. 

When both FSS and laser-induced splitting are present, the laser polarizations have a slightly different impact. The difference between the concurrence obtained for horizontal and diagonal polarization is estimated as
\begin{equation}
\Delta C \approx f(\gamma\sbt{B},\Delta_\text{FWHM}) \left[ 1-C_0(\gamma\sbt{X},\delta) \right]
\end{equation}
and increases with pulse duration and FSS. This implies an optimal laser polarization can be found. By analyzing Eq.~\eqref{eq:C_full_estimate}, one finds that the horizontal polarization corresponds to the optimal choice, while diagonal polarization is the most detrimental one. 

Additionally, we have performed numerical calculations with a rectangular pulse with smoothened edges (cf., Supplemental Material \cite{supp}). The resulting concurrence shows only negligible differences to the case of a Gaussian pulse. Note that the analytic expressions do not depend on the pulse shape and that the effect stems from a symmetry breaking related to the laser polarization. This underlines that the pulse shape only marginally influences the upper bound for the concurrence.

In conclusion, our analysis shows that the TPE scheme with a pulse of finite duration sets a limitation for the degree of entanglement of photon pairs due to the excitation itself. This result is supported by numerical calculations for a two-photon $\pi$ pulse as well as analytical expressions. Its generic nature is explained by an intuitive physical picture of a Stark-induced energy splitting, which introduces which-path information, and thus, reduces the entanglement. The effect increases for longer pulses, and, in principle, disappears in the limit of instantaneous excitation. However, in this limit the excitation model and the TPE scheme become inadequate. In most practical situations, a pulse shorter than 2-3~ps produces unwanted exciton states in typical InGaAa or GaAs QDs. Consequently, while the FSS in QDs can be reduced to zero, the pulse length cannot, and this sets an upper limit for the on-demand generation of entangled photon pairs. Note that we also expect the indistinguishability of photons from subsequent TPE pulses to be limited due to the dynamic laser-induced energy shift.

Our calculations accounted only for the most basic relaxation mechanisms that are present in all realizations of a four-level emitter, i.e., rates for the cascaded decay. Further influences that might affect the entanglement such as, e.g., phonons were disregarded. Our analysis thus explores an ideal situation highlighting the detrimental effect of the excitation scheme even in the absence of other destructive mechanisms.

Finally, we note that our theoretical prediction of $C\approx 0.975$ is very close to the highest experimentally achieved concurrence of 0.97(1)\cite{Huber_PRL_2018}. This may indicate that the laser-induced which-path information provides a quantitative explanation for the observed deviation from unity.

\section*{Acknowledgments}
T. K. B. and D. E. R. acknowledge support by the Deutsche Forschungsgemeinschaft (DFG) via the Project No.~428026575.
A. R. acknowledges the Austrian Science Fund (FWF) via the Research Group FG5 and the I 4380, European Union’s Horizon 2020 research and innovation program under Grant Agreements No.~899814 (Qurope) and No.~871130 (Ascent+). This project was funded within the QuantERA II Programme that has received funding from the European Union’s Horizon 2020 research and innovation programme under Grant Agreement No.~101017733, and with funding organization the Austrian Research Promotion Agency (FFG), Grant No.~891366.
We are further grateful for support by the Deutsche
Forschungsgemeinschaft (DFG, German Research Foundation) via Project No.~419036043.

\bibliography{PIbib}

\end{document}